\documentclass[prb,twocolumn,floatfix,showpacs,superscriptaddress]{revtex4}
\usepackage{graphics}
\usepackage{epsfig}
\usepackage{times}

\begin{document}
\preprint{HEP/123-qed}
\title{Density-matrix renormalization study of the frustrated fermions on the triangular lattice}
\author{Satoshi Nishimoto}
\affiliation{Leibniz-Institut f\"ur Festk\"orper- und Werkstoffforschung Dresden, D-01171 Dresden, Germany}
\author{Chisa Hotta}
\affiliation{Kyoto Sangyo University, Department of Physics, Faculty of Science, Kyoto 603-8555, Japan} 
\date{\today}
\begin{abstract}
We show that the two-dimensional density-matrix renormalization analysis is useful to detect the symmetry 
breaking in the fermionic model on a triangular lattice. 
Under the cylindrical boundary conditions with chemical potentials on edge sites, 
we find that the open edges work as perturbation to select the strongest correlations 
{\it only in the presence of a long range order}. 
We also demonstrate that the ordinary size scaling analysis on the charge gap as well as that 
of the local charge density under this boundary condition could determine the metal-insulator phase boundary, 
which scales almost perfectly with the density of states and the exact solutions 
in the weak and strong coupling region, respectively. 
\end{abstract}
\maketitle
\narrowtext 
%
Geometrically frustrated lattices sometimes provide particular 
situation where the strong correlation between particles or spins 
works effectively to give rise to exotic states. 
The numerical approach to such systems made remarkable progress in recent years; 
quantum monte carlo method (QMC) is applied to clarify the supersolid state 
on a triangular and square lattices\cite{wessel05,mila08}, plaquette state on a quantum dimer model, 
the nematic orders in the ring exchanged ladders\cite{lauhili05}. 
Also the exact diagonalization by the symmetry analysis 
gave much information on the low-energy excitation of the Kagome lattice\cite{misguish02}, 
nematic orders in the frustrated square lattice\cite{nic06}, and so on. 
In the present paper we apply the density matrix renormalization group (DMRG) 
method to the two-dimensional(2D) frustrated system. 
So far, we understand that the systematic treatment to 2D-DMRG is not known 
to such system where a competition between several orders exists. 
We give a prototype analysis on how to determine the translational symmetry breaking, 
to detect the metal-insulator transition and so on. 
We choose as a representative frustrated system a fermionic system on the triangular lattice. 
Since the QMC is quite hopeless to such system due to the minus sign problem, 
the 2D-DMRG approach is useful instead, 
which can cope with much bigger sizes compared to the exact diagonalization. 
However, it is relatively quite difficult to get a reasonable results in DMRG when the system 
is critical or in higher dimension than one, 
which is demonstrated in the $t$-$J$ model on a square lattice\cite{white04}. 
Recent 2D-DMRG analysis with the cylindrical boundary condition is given 
by systematically varying the aspect ratio of the finite cluster, 
which successfully determined the magnetization of the 
N\'eel order in the square lattice Heisenberg model\cite{white07}.  
However, the same analysis for the triangular lattice Heisenberg model turned out to be 
far difficult. 
In the present paper, we adopt the same cylindrical boundary condition, 
and demonstrate that the triangular lattice geometry combined with this boundary condition 
is useful to detect the symmetry breaking in an unbiased manner, 
which is also possible in the case of other lattices, e.g., the bipartite square lattice. 

We mention in advance that even though the present analysis turned out to be 
successful, there are only few cases where the 2D-DMRG is reasonably adopted; 
(1) in detecting the Ising type of long range orders (with quantum fluctuation) 
when the ground state candidates are elucidated, 
(2) when the numerous number of states (more than the order of $\sim 10000$) 
are kept by e.g., using the parallel computing system\cite{hager05}, 
(3) analysis on metallic state away from the phase boundary. 
Still, the difficulty depends on what quantity we measure, 
and the numerically rigorous data provided by the DMRG would be of help in many occasions 
as in the present case which is one of the examples of case (1). 
\par
\begin{figure}[tbp]
\begin{center}
\includegraphics[width=6.5cm]{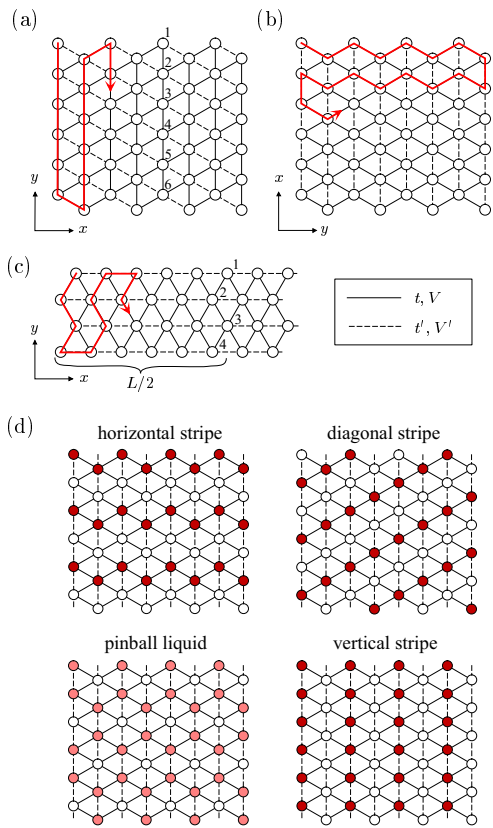}
\end{center}
\caption{(Color online) 
(a)--(c) Schematic representation of the finite cluster, where we take $x$- and 
$y$-direction as OBC and PBC, respectively. The block states are 
constructed along the one-dimensional chain array shown in arrows. 
The chemical potentials, $\mu$, are placed on each open edge sites. 
In (a)-(c) the dotted lines which are anisotropic bonds with $V_{ij}=V'$ and 
$t_{ij}=t'$ are taken in different directions. 
(d) Schematic configurations of particles on the cluster in different phases 
of the $t$-$V$ model; (i) horizontal or diagonal stripes for $V'>V$, 
vertical stripe for $V'<V$, and pinball liquid for $V'\sim V \gtrsim 3t$.
}
\label{f1}
\end{figure}

We choose as the simplest fermionic model the $t$-$V$ model whose Hamiltonian given as, 
\begin{equation}
{\cal H}_{t-V} =\sum_{\langle i,j \rangle} t_{ij}\left(c^\dagger_i c^{\vphantom{\dagger}}_j + {\rm H.c.}\right) 
+  \sum_{\langle i,j \rangle}V_{ij} n_i n_j. 
\label{tvham}
\end{equation}
Here $c^{\dagger}_j$ ($c^{\vphantom{\dagger}}_j$) are creation (annihilation) operators 
of fermions and $n_j$ (=$c^\dagger_j c^{\vphantom{\dagger}}_j$) are number operators. 
The interactions act only between neighboring sites $\langle ij \rangle$. 
Anisotropies of the hopping amplitudes and repulsion strengths are all positive and are given by 
$(t_{ij},V_{ij})=(t',V')$ for the vertical bond 
and $(t,V)$ for the remaining bond directions, as shown in Fig.~\ref{f1}(a)--(c). 
We focus on half-filling where we have competitive orders due to 
the commensulability of charges, namely one charge per two sites. 
We take $t=1$ as the unit of energy and fix $t'=1$ if not otherwise stated.
\par
The overall ground state phase diagram of the $t$-$V$ model is presented by part of the authors based on 
the exact diagonalization on a $4\times 6$ cluster and on a strong coupling analysis\cite{hotta06-2}. 
When the interaction is enough large, the phase diagram is characterized by the three different 
phases according to the anisotropy of $V$ and $V'$ as shown schematically in Fig.~\ref{f1}(d). 
Around the regular triangular geometry of the interactions, $V'\sim V$, 
we have a partially charge ordered liquid called a ``pinball liquid". 
This phase breaks the translational symmetry as characterized by the wave vector number, 
${\bf k}=\pm (\frac{2\pi}{3},\frac{2\pi}{3})$, 
and it originates from the long range order of "pins", while still retains a coherent metallic property. 
If one enters the anisotropic region, $V'>V$, one finds an insulator where the particles align in stripes 
in the horizontal (or diagonal) direction in order to avoid the energy rise by the stronger interaction, 
$V'$. We also have similar stripes in $V'<V$ which extends along the vertical direction. 
The weakly coupled to intermediate coupling region is, however, not clarified yet, 
which will be focused in the present paper. 
\par
%
\par
The 2D-DMRG calculation is performed on a $N=N_x\times N_y$ cluster shown in Fig.~\ref{f1}(a)--(c). 
We keep up to 3200 basis for each DMRG block and undergo $\sim 20$ sweeps until 
the ground-state energy converges within an error of $\sim 10^{-10}t$. 
Here, we adopt the open (OBC) and periodic (PBC) boundary condition in $x$- and $y$-direction, respectively as shown schematically in Figs.~\ref{f1}(a)--(c). 
It is well known that if all the boundaries are taken as periodic, 
the translational symmetry of charge density is always preserved at finite system size 
since the wave function is the superposition of all the degenerate states with equal weight. 
However, if we take the the cylindrical topology of the triangular lattice system and opens 
part of the boundaries, it gives rise to the lifting of the degenerate wave functions 
in the DMRG calculation. Therefore, we can easily detect translational-symmetry breaking 
charge-ordered states shown in Fig.~\ref{f1}(d). We additionally place the chemical potentials, 
$\mu=(V+V')/2$ or $V$ on each open edge site in order to suppress the artificial Friedel 
oscillation induced by OBC. 

%
\begin{figure}[tbp]
\begin{center}
\includegraphics[width=8.5cm]{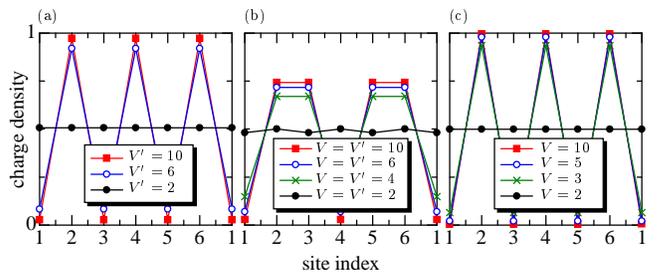}
\end{center}
\caption{(Color online) Charge density at sites indexed by numbers 1 to 6 in the cluster Fig.1(a), 
where the parameters are taken as (a) $V=1$, (b) $V=V'$, and (c)$V'=1$ 
under the variation of $V$ or $V'$. 
The calculations are given on $8 \times 6$ cluster (Fig.1(a)) with periodic boundary in the $y$-direction 
and open boundary in the $x$-directions. 
}
\label{f2}
\end{figure}

\par
Usually, in the exact diagonalization and other finite size methods, 
we figure out what kind of structural factor the state has by analyzing the two-point correlation functions. 
If the considered structure has a true long range order, the amplitude of the structural factor 
shall remain finite in the thermodynamic limit after the finite scaling analysis. 
However, in the present 2D-DMRG calculation, each of these phases are detected more simply 
by the spatial structure of the charge density as follows; 
as in Fig.~\ref{f1}(a) we take the PBC on one of the $V$-bond directions. 
Then, the other two open bonds with different interactions, $V$ and $V'$, 
artificially lifts the degeneracy of the finite size ground state and enables us to distinguish several 
translational symmetry broken and also the unbroken phases. 
Figure~\ref{f2} shows the charge density of each characteristic state 
for sites indexed by numbers 1 to 6  in Fig.~\ref{f1}(a). 
The two-fold periodic structure in Fig.~\ref{f2}(a) and (c) indicate the diagonal 
and vertical stripes, respectively. 
We also detect the three fold periodic rich-rich-poor density of charges (denoted as A-A-B structure) 
in the pinball liquid state in Fig.~\ref{f2}(b). 
In contrast to those phases, the metallic phase at small $V$ or $V'$ has almost 
uniform charge density along this periodic boundary. 
Therefore, the way we assign the boundary conditions should be the one that could 
efficiently detect the differences between ordered states. 

\par
Here, we must note that the horizontal stripe, which is another possible 
two-fold periodic structure at $V'>V$, 
is not realized if we take the boundary conditions as in Fig.~\ref{f1}(a). 
Therefore, we consider another type of cluster shown in Fig.~\ref{f1}(b) 
where both the diagonal and horizontal stripes are compatible, 
and compare their energies. 
The calculations on the $6 \times 8$ cluster confirm 
that the horizontal order is always the ground state in the insulating phase 
at at $V'>V$ 
and the diagonal order belongs to a higher state with small excitation energy. 
%
\begin{figure}[tbp]
\begin{center}
\includegraphics[width=8.0cm]{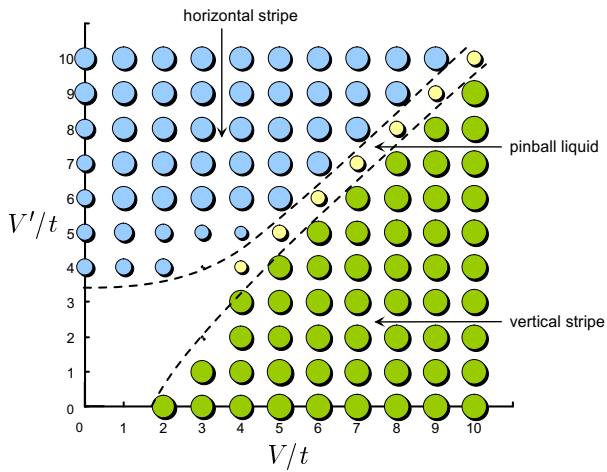}
\end{center}
\caption{(Color online) Phase diagram of the $t$-$V$ model on the plane of $V/t$ and 
$V'/t$ obtained by the DMRG calculation. 
The size of the circle corresponds to the amplitude of the charge density modulation. 
The calculations are given on $8 \times 6$ (or $6 \times 8$) cluster in Fig.~\ref{f1}. 
}
\label{f3}
\end{figure}

\par
We now present the phase diagram on the plane of $V'$ and $V$ in Fig.~\ref{f3}. 
As mentioned earlier we find three characteristic phases in the strong interaction region; 
the pinball liquid phase is sandwiched between two different stripes. 
As has been discussed in Ref.[\onlinecite{hotta06,hotta06-2}], the width of the pinball liquid phase is 
determined by the absolute value of $t$ and $t'$, e.g. $\frac{3t}{2}$ for $t<t'$. 
Since these three phases breaks the translational symmetry with different 
characteristic wave numbers the transition between them are of first order. 
In the strongly interacting region, the phase boundaries show excellent agreement with the 
one found in the exact diagonalization. The similar phase diagram has been obtained 
by the same 2D-DMRG method for $t$-$U$-$V$ model~\cite{nishimoto08}, and 
by the variational monte carlo (VMC) method~\cite{watanabe06}. 
The physics of the charge ordering is essentially the same but the pinball-liquid phase is replaced by 
the three-fold charge-ordered metallic phase with doubly occupied sites.
\par
In one dimension, the open boundary is regarded as impurities which induces a Friedel oscillation 
in metals and can be used to analyze the wave number and critical exponent of 
the Tomonaga-Luttinger liquid\cite{shibata96,affleck02}. 
These Friedel oscillations are not a true long range order and shall be suppressed if 
the proper chemical potentials are placed on open edges. 
On the other hand, if the system has a true long range order with particular symmetry $k\ne 0$, 
the OBC actually lifts the degeneracy of wave functions, 
regardless of whether the chemical potentials is present or not. 
We take advantage of this fact in 2D and use the partially open edges 
which would give rise to the spatial structure of the local quantities (charge density) 
{\it only when we have long range order}. 
The partially opened edge must be compatible in geometry with the symmetry of the orders 
to get rid of the the translated counterpart components. 
We note that the chemical potentials not only suppresses the Friedel oscillation but 
also helps the variational states not to fall into the local minimum in the case of the 
Ising-type of orders. 
Even if we adopt this method to the square lattice system, such breaking of symmetry will also appear. 
This is because the translational symmetry is broken in a direction with open ends 
and concomitantly the breaking of symmetry in the other direction which is 
periodic occurs. 
From these results, we may argue that the analysis based on the cylindrical boundary condition 
are useful to detect the breaking of translational symmetry in the two-dimensional 
lattices. 
We also confirmed that the same analysis is well adopted to the the Kagome lattice as well
\cite{nishimoto}. 
\par
%
\begin{figure}[tbp]
\begin{center}
\includegraphics[width=8.5cm]{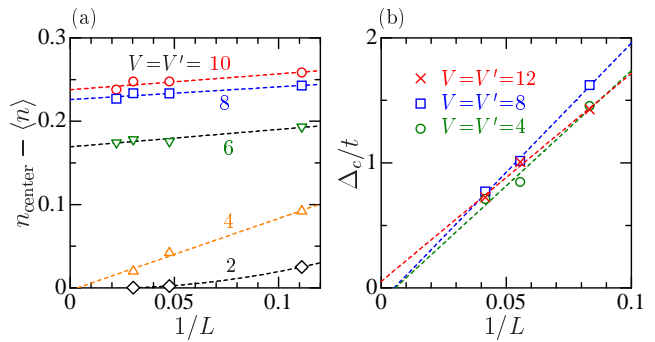}
\end{center}
\caption{(Color online) Finite size scaling of the expectation value of 
(a) charge density extracting the mean charge density at the center four sites 
of the system, and (b) the charge gap at $V=V'$ and $t=t'=1$.
The calculations are given on $4 \times L$ cluster with periodic boundaries in the $x$-directions 
and open boundary in the $y$-direction. 
}
\label{f4}
\end{figure}

In the above analysis, we only considered the fixed cluster size. 
However, the phase boundary is usually overestimated by the small system size, 
particularly when we have finite but long correlation length compared to the size of the cluster. 
Thus one cannot really discriminate the short range order from the true long range order 
without the proper size scaling analysis. 
In order to get the more quantitative information, we give the scaling analysis on these 
local quantities by using the $L\times 4$ cluster shown in Fig.~\ref{f1}(c). 
We focus on the four different sites placed at the center of the cluster 
(see the sites marked by numbers 1--4 in Fig.~\ref{f1}(c)) which are the least influenced 
by the open boundary. 
Figure~\ref{f4}(a) shows the deviation from the average charge density of these sites 
as a function of inverse system length $L^{-1}$ in the pinball liquid phase at $V=V'$. 
Here we take $L=6n$ with integer $n$, which is a period compatible with both the two-fold and 
three-fold periodicities. 
As we can see, at $V=V'<6$  the deviation of charge density is extrapolated to zero, 
while over $V'=V\ge 6$ they change the slope significantly and become finite. 
This means that the translational symmetry is broken at $V=V'\sim 6$, and a three-fold periodic 
long range order appears. 
This phase boundary is smaller than the one estimated from 
the VMC simulations on the same model, $V_c\sim 12t$\cite{miyazaki09}. 
The VMC estimation based on the Fermi sea slater wave function is an upper bound, 
and the present results lie just in between the Hartree-Fock estimation and the VMC one. 
\par
We also estimate the charge gap, $\Delta_c$, of the three-fold periodic phase. 
The charge gap at each system size $N$ and particle number $N_e=N/2$ is given as, 
$\Delta(N_e,N_x,N_y)= E(N_e+1,N_x,N_y)+E(N_e-1,N_x,N_y)-2E(N_e,N_x,N_y)$, 
where $E(N_e,N_x,N_y)$ is the energy of the corresponding cluster. 
Figure~\ref{f4}(b) shows the extrapolation of 
$\Delta(N_e,N_x,N_y)$ towards $N_x=L, L\rightarrow \infty$ with $N_y=4$. 
We see that the system can be regarded as gapless in the thermodynamic limit, 
which is consistent with the previous study\cite{hotta06}. 
The estimated phase boundary quantitatively agrees with the one 
by the charge density in Fig.~\ref{f2} and guarantees the present analysis 
on the local quantities. 

\begin{figure}[tbp]
\begin{center}
\includegraphics[width=5.5cm]{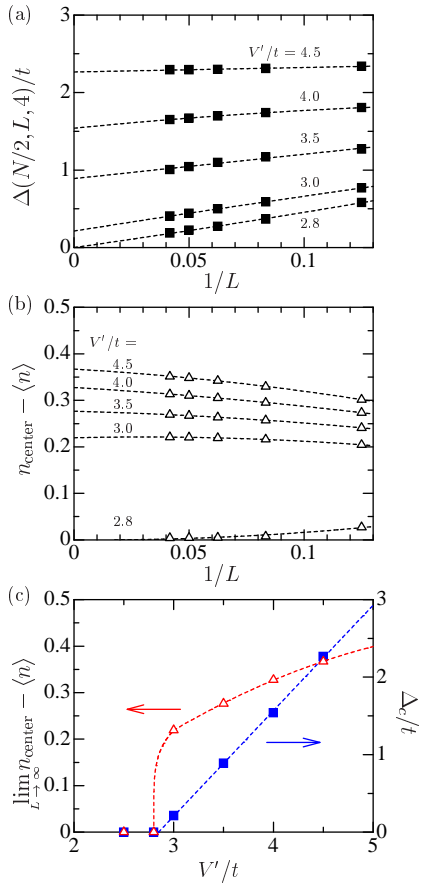}
\end{center}
\caption{(Color online) 
Finite size scaling of the expectation value of 
(a) charge density after extracting the mean charge density at the center four sites 
of the system, and (b) the charge gap at $V=0$ and $t=t'=1$. 
In panel (c), the gap and the charge density disproportionation 
in the bulk limit is presented as a function of $V'/t$. 
}
\label{f5}
\end{figure}
%
\begin{figure}[tbp]
\begin{center}
\includegraphics[width=5.5cm]{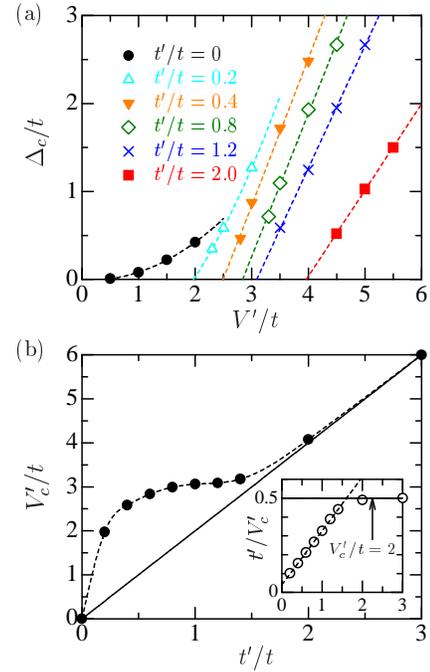}
\end{center}
\caption{(Color online) 
(a) The charge gap in the bulk limit 
at $V=0$ as a function of $V'$ for the several choices of $t'/t$. 
The finite size scaling is given in the same way as those in Fig.4. 
(b) $t'/t$-dependence of the phase boundary between the normal weak coupling Fermi liquid phase 
and the horizontal stripe insulator, $V_{\rm MI}$. 
The inset shows the $t'/t$-dependence of $t'/V_{\rm MI}$ for the same data. 
}
\label{f6}
\end{figure}

\par
Now we focus on the weak coupling region of the phase diagram which is not well studied yet. 
In the horizontal stripe phase at $V'>V$, the charge gap opens due to the staggered 
long range order of particles in the anisotropic direction. 
The gap is well estimated by extrapolating the system size in that direction. 
Figure~\ref{f5}(a) shows the $L^{-1}$-dependence of the gap for 
several choices of $V'/t$ with the fixed $V=0$. 
The same scaling of the center site charge density is given in Fig.~\ref{f5}(b). 
As plotted in Fig.~\ref{f5}(c), the smooth opening of the extrapolated gap 
suggests the second order transition, while the extrapolated charge density shows 
a sharp development at the transition point. These are common behaviors in 
charge ordering induced by long-range interactions~\cite{nishimoto03}. 
However, the transition point itself shows an excellent agreement between these two quantities. 
We note that the $L$-dependence of the gap is very small in the insulating state, 
which reflects the localized character of the wave function. 
Resultantly, although the phase boundary of Fig.~\ref{f3} shifts somewhat to a smaller 
$V$- or $V'$-region after the scaling, the correction is less than $\sim 0.5t$. 
\par
Next, by performing the same analysis, 
we investigate the effect of varying the geometry of transfer integrals, $t'/t$. 
Figure~\ref{f6}(a) shows the extrapolated charge gap for several choices of $t'/t$. 
By fitting the $V'$-dependences, the metal-insulator (MI) phase boundary is detected. 
Here, the gap opens linearly which makes the estimate precise. 
The $t'/t$-dependence of $V'_{\rm MI}$ is plotted in Fig.~\ref{f6}(b).  
Just off the non-interacting point, $V=0, V'\sim 0$, the gap opens since the shape of 
the Fermi surface at half-filling is a regular square and the perfect nesting takes place. 
Therefore the obtained insulator is a typical charge-density-wave state with a small gap. 
The value of $V'_{\rm MI}$ significantly changes at $0\le t'/t \le 0.5$, 
and then saturates to $V'_{\rm MI} \sim 3$. 
At $t'/t \gg 1$, the system is regarded as the one-dimensional chain, 
where the exact solution of the charge gap is given as $V'_{\rm MI}/t'=2$\cite{hubbardexact}. 
This can be more easily understood in the $V'_{\rm MI}/t$-$t'/t$ plot in the inset, 
where the value of $V_{\rm MI}'/t'$ shows a clear crossover from the $V'_{\rm MI}$-linear term 
to the constant value at around $V'_{\rm MI} \sim 3$. 
\par

%
\begin{figure}[tbp]
\begin{center}
\includegraphics[width=5.5cm]{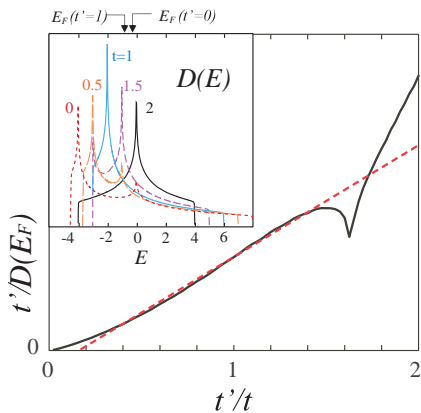}
\end{center}
\caption{(Color online) $t'/D(E_F)$ as a function of $t'/t$ at the fixed value of $t=1$, 
where $D(E_F)$ is the non-interacting density of states of the anisotropic 
triangular lattice at the Fermi level at half-filling. 
The inset shows the non-interacting density of states for several choices of $t'/t$. 
}
\label{f7}
\end{figure}

We now try to understand the characteristic $t'/t$-linear behavior of $V_{\rm MI}'/t'$ 
in the weak coupling region. 
Figure~\ref{f7} shows the $t'/t$-dependence of 
the non-interacting density of states of the anisotropic triangular lattice 
at the Fermi level at half-filling, $D(E_F)$. 
We find that $t'/D(E_F)$ scales almost linearly with $t'/t$ in the wide region, 
$0.25\le t'/t \le 1.5$, 
which is exactly the region where we find the same linear relation in the inset of Fig.~\ref{f6}(b). 
By interpolating these two characteristic relations, we obtain the following relation, 
\[
\frac{V_c'}{t'}= \frac{1}{\alpha+\beta t'/D(E_F)}, 
\]
where $\alpha=0.04, \beta=0.06$. 
The two different linear relations found in Fig.~\ref{f5}(c) indicates that the 
MI transition changes its character from the weak to the strong coupling ones, 
i.e. from the charge-density-wave to the charge ordered insulator, 
at $t'/t\simeq 1.6$, 
which is also the point where the van Hove singularity touches the Fermi level at half-filling. 
\par
To summarize we performed the 2D-DMRG analysis on the fermionic system 
on an anisotropic triangular lattice. 
We showed that in the triangular lattice the cylindrical boundary condition with 
chemical potentials on open edge site, the translational symmetry breaking is detected 
by measuring the local quantity in a reasonable accuracy. 
Also the finite scaling analysis is found to give a further quantitative estimation 
on the phase boundaries. 
The demonstration is given on the determination of the metal-insulator phase boundary 
as a function of the degree of the transfer integral, $t'/t$. 
It is found that at relatively small $t'$ the phase boundary is scaled by the non-interacting 
density of states at the Fermi level, which indicates that the phase transition is 
regarded as a charge-density-wave transition due to the instability if the Fermi surface. 
On the other hand, at $t'\ge 1.6t$, the phase boundary is scaled to $V_c'=2t'$, which means 
that the system undergoes the phase transition to the charge ordered state in the one-dimensional manner. 
\par
The role of $t'/t$ on the metal-insulator transition has been discussed previously 
in some articles related to organic compounds based on the mean-field approach\cite{chisa03}. 
Recently, it is proved by the uniaxial strain experiment that 
the MI transition temperature of the $\theta$-ET$_2$CsZn(SCN)$_4$ into the 
horizontal charge ordered state significantly increases when the geometry of the 
transfer integral varies as $t'/t=0 - 0.5$\cite{kondo06}.  
The tendency that the insulator is stabilized due to $t'/t$ is contrary to what we get in the 
present analysis. This might be because the present model deals with the half-filled charge density 
wave, while the experiments should be better explained in the 3/4-filled extended Hubbard model 
(which has a different Fermi level and a Fermi surface) instead. 
We expect that even in this latter model the $V_{\rm MI}$ will be scaled by the density of states. 
The present paper shows for the first time that the 2D-DMRG provides a reliable way to cope with 
the difficult parameter region where the electronic correlation and kinetic energy compete with 
each other, and presented the clear crossover from the weak to strong coupling. 


\end{document}